\documentclass[aps,pre,twocolumn,superscriptaddress,nofootinbib]{revtex4-2}

\usepackage{amsmath,amssymb}
\usepackage{graphicx}
\usepackage{xcolor}
\usepackage{hyperref}

\newcommand{\W}{W_{ij}}

\begin{document}

\title{Morphology of frozen labyrinths from irreversible threshold dynamics}

\author{Daniel Richard Levy}
\affiliation{Program in Applied and Computational Mathematics, Princeton University, Princeton, NJ 08540, USA}

\date{\today}

\begin{abstract}
Majority threshold dynamics, in which each agent adopts the dominant state in a
weighted neighborhood, relaxes a binary field toward consensus or stripes. We study what happens when this rule is made irreversible: each agent, interacting through a
Gaussian kernel on a lattice, may flip out of its local weighted minority at most
once. The reversible form is threshold dynamics of Merriman--Bence--Osher type, an
exactly solvable calibration in which interfaces move by mean curvature with
closed-form lattice pinning and mobility. Irreversibility changes the outcome.
From random initial conditions, the one-flip rule freezes balanced non-consensus
labyrinths that reversible relaxation drives away. The patterned regime is a
window of initial spin compositions around equal balance, narrowing as the
interaction range grows, controlled by a standardized bias whose onset is
independent of scale over a fourfold range. The frozen morphology is arrested
coarsening: bicontinuous at balance, with a feature width that grows sublinearly
and falls below the interaction range at large scales. We determine the mechanism
by intervention. Fixing the initial condition while varying the update order shows
that the coarse domain layout is deterministic, while the update order enters only
in a secondary first-passage race that fine-tunes the wall positions the
deterministic dynamics has already set. Those walls are marked by frustration:
agents frozen against their own local field. This signature is identically absent
from any reversible relaxation, is overwhelmingly interfacial, and carries almost
none of the pattern's large-scale shape.
\end{abstract}

\maketitle

\section{Introduction}
\label{sec:intro}

Local alignment erases micro-level structure. When each site of a spatially extended system
repeatedly adopts whichever state its neighborhood favors, small domains are
absorbed into larger ones, interfaces straighten, and the configuration
coarsens toward consensus or stripes. The detail present in the initial condition is
progressively lost. Threshold dynamics is the minimal rule with this behavior:
replace the state of each site by the sign of a Gaussian-weighted average of
the states around it. In the limit of vanishing kernel width the interface
between the two phases moves with velocity equal to its mean curvature---the
gradient flow that shortens it---which is the content of the
Merriman--Bence--Osher (MBO) scheme \cite{merriman_motion_1994}. 

On a lattice the erasure is incomplete. A single step displaces the interface
by an amount proportional to its curvature, and when that displacement falls
below one lattice spacing, the interface stops moving altogether. Therefore, flat and weakly curved
interfaces are \emph{pinned} \cite{merriman_motion_1994}. This pinning is the threshold-dynamics
counterpart of propagation failure in discrete reaction--diffusion systems
\cite{keener_propagation_1987}. The same competition between curvature drive
and lattice discreteness organizes zero-temperature kinetic Ising dynamics
\cite{spirin_freezing_2001} and its $Q$-state generalization in Potts grain growth \cite{holm_effects_1991}, where curvature-driven interfaces freeze once their motion drops below the lattice scale. 

Discreteness is one limit on erasure; irreversibility is another, and the latter is the
subject of this paper. All of the dynamics above are reversible at the level of
the individual site: a site that switches one way may switch back later, and
the system keeps relaxing until nothing wants to move. In this paper, we ask what happens when
each site may change state only \emph{once}. The motivation is that many
collective processes are built from commitments rather than adjustments. An
agent who has adopted, joined, or taken a side often does not revert on the
timescale over which the process unfolds
\cite{watts_simple_2002,gleeson_seed_2007,centola_complex_2007}. Such systems pose a
standing puzzle: local imitation drives toward consensus, yet real
committed-choice processes (e.g., competing contagions, competing technology or
crystal domains, invasion fronts in ecology) routinely freeze into persistent
mixed states instead. We show that permanent, two-sided commitment in threshold dynamics freezes
balanced non-consensus labyrinth patterns generically. We study the geometry of
these patterns and what drives it: the coarse domains are laid down by the
initial condition and the deterministic dynamics together, their walls are refined mostly by the deterministic advance of the competing
fronts, and the frozen state carries a measurable interfacial signature that no
reversible relaxation leaves. Where the walls of such a two-sided invasion end up,
and the signature they carry, are concrete transferable observables.

Irreversibility enters this literature in several inequivalent forms. Zealots
and committed agents \cite{galam_rational_1997,mobilia_role_2007} form a quenched subpopulation,
fixed in advance and never updated. The question there is whether their
presence blocks consensus, while the remaining agents relax freely. In the
persistent voter model \cite{latoski_opinion_2024}, accumulated confidence renders agents
temporarily immovable, generating an effective surface tension and
curvature-driven coarsening, but here the immobility decays and the system still
orders. Bootstrap percolation and threshold cascades
\cite{aizenman_metastability_1988,holroyd_sharp_2003,watts_simple_2002} are permanently irreversible
and also \emph{monotone}: a single state spreads, the occupied set only grows,
and no rival front is ever met. Irreversibility that is both permanent and two-sided
is the remaining case: both phases invade, they collide, and neither can
reclaim what it loses. Its defining fingerprint is an observable no reversible
relaxation possesses. An energy-lowering flip is taken when it does not
raise the energy, so a reversible dynamics can never come to rest with a spin
strictly against its local field. But the one-flip rule removes this guarantee, and
the absorbing states carry a finite density of \emph{frustrated} spins---frozen
against their field---which we use throughout as the signature of
irreversibility.

The model is minimal. Spins $s_i\in\{\pm1\}$ sit on $\mathbb{Z}^2$ and interact
through an isotropic Gaussian kernel of range $\sigma$; a spin strictly in its
local weighted minority flips at rate $\beta$; each spin flips at most once.
Deleting the last clause recovers a reversible threshold dynamics of MBO type,
so the one-flip rule is the \emph{only} difference between the two dynamics we
compare, and any difference in outcome is attributable to it. The reversible
side serves as calibration and we treat it exactly where we can: flat
interfaces pin with a closed-form constant, curved interfaces erode with the
continuum curvature drive, and droplets have a crossover radius
$R_c\sim\sigma^2$ (Secs.~\ref{sec:flat} and \ref{sec:droplet}). The
irreversible side is where the new behavior lies.

Three results follow. First, starting from a random mixture in which each site
is $+1$ with probability $p_0$, there exists a window of compositions around
$p_0=\tfrac12$ inside which the system freezes a balanced two-phase pattern
rather than being swept to consensus. What controls the window is the initial bias measured against the fluctuations a site actually feels, $z_0\simeq2\sqrt{\pi}(2p_0-1)\sigma$. The window width is proportional to $1/\sigma$, since the boundary sits at a value of $z_0$ that does not drift with $\sigma$
(Sec.~\ref{sec:window}). That boundary appears to be dynamical rather than geometric: the
minority survives past every static percolation threshold of the initial field
we examined. 

Second, the frozen pattern at balance is a smooth bicontinuous labyrinth with no
preferred wavelength (unlike a Turing pattern), best read as coarsening caught in the act and held there. Its characteristic width grows with $\sigma$ but sublinearly, so the scale of the frozen patterns is not simply the
interaction range (Sec.~\ref{sec:morph}). 

Third, we determine the causal structure of the frozen pattern by intervention. Holding the initial condition fixed while varying the update order separates what is deterministic from what depends on the update order, and ablating the frustrated sites separates what places the walls from what merely
locks them. The coarse domains are deterministic---laid down by the initial
condition and the dynamics together. The positions of the domain walls
are refined predominantly by the deterministic advance of the
competing fronts, with a secondary first-passage race---the sole role of the
update order---fixing only their last fraction of $\sigma$; this race has a
counterpart in the competition interfaces of first-passage percolation
\cite{haggstrom_first_1998}. Frustration marks the seams where the fronts
meet but, as ablation shows, does not place the walls: it locks them, with any isolated frustrated sites being residue of the collision rather than its cause. 

\section{Model}
\label{sec:model}

In our model, the sites of $\mathbb{Z}^2$ carry spins $s_i\in\{-1,+1\}$. Sites interact through a
Gaussian kernel with the self-interaction removed,
\begin{equation}
W_{ij}=\exp\!\left(-\frac{\|x_i-x_j\|^2}{2\sigma^2}\right),
\qquad W_{ii}\equiv0,
\label{eq:kernel}
\end{equation}
so that each site responds to its surrounding neighborhood rather than to its
own state. Each site feels the local weighted field
\begin{equation}
m_i=\sum_{j\neq i}W_{ij}\,s_j ,
\label{eq:field}
\end{equation}
and is \emph{eligible} to flip when it lies strictly in its local minority,
$s_im_i<0$. In the \textbf{irreversible (one-flip) model}, each eligible,
not-yet-flipped site flips $s_i\to-s_i$ at rate $\beta$ (which sets only the
timescale; we take $\beta=1$) and thereafter never flips again. We take the dynamics to be asynchronous and continuous in time: sites act on independent clocks. The state $(\mathbf{s},\mathcal U)$, with
$\mathcal U$ the set of sites that have not yet flipped, is Markovian. The
eligible set $\{i\in\mathcal U:s_im_i<0\}$ and the exponential clocks are
determined by the state alone. Since $|\mathcal U|$ strictly decreases at every
event, the process is absorbed within $L^2$ events. The \textbf{reversible
variant} drops the one-flip constraint: eligible sites flip whenever
$s_im_i<0$.

Both the irrerversible model and its reversible variant flip only sites with $s_im_i<0$. Because $\W>0$ for every pair,
\begin{equation}
H=-\tfrac12\sum_{i\neq j}\W\,s_is_j
\label{eq:energy}
\end{equation}
is a ferromagnetic energy: alignment is favored at all separations and the
ground states are the two consensus states. Isolating the terms containing
$s_i$ gives $H=-s_im_i+\text{const}$, and since $m_i$ excludes $s_i$ it is
unchanged when $i$ flips, so a single flip changes the energy by
$\Delta H=2s_im_i<0$. Asynchrony is essential to the monotonicity of $H$: when two neighbors flip
simultaneously the energy change acquires an indefinite cross term
$\Delta H=2s_im_i+2s_jm_j-4\W s_is_j$, whose last contribution is positive for
anti-aligned pairs. So strongly coupled, marginally eligible neighbors flipping
together can raise $H$ and synchronous threshold dynamics need not descend
monotonically. 

The irreversible model and its reversible variant therefore descend the same
landscape and differ only in where they stop. In the reversible model a site
is eligible whenever $s_im_i<0$, so the dynamics can come to rest only when
$s_im_i\ge0$ for every $i$: it halts precisely at configurations stable to
single-spin flips. The irreversible model may halt earlier,
frozen in a configuration in which some \emph{spent} sites still satisfy
$s_im_i<0$. These sites, which we call \emph{frustrated}, index the
single-flip moves that would lower $H$ but that the dynamics can no longer
execute. They are the central observable of Sec.~\ref{sec:mechanism}.

The update rule---align each site with the sign of the Gaussian-convolved
field---is that of two-phase threshold dynamics, and in synchronous reversible
form it is the MBO scheme \cite{merriman_motion_1994} up to the self-exclusion $W_{ii}=0$,
which is what makes the lattice flat interface pinned rather than marginal
(Sec.~\ref{sec:flat}). We use the asynchronous relaxation of the same rule. The
correspondence calibrates the single-interface sector against established
results: the content of Secs.~\ref{sec:flat} and \ref{sec:droplet} lies in the
exact lattice constants, not in the phenomena. It also delineates what is new,
since threshold dynamics of MBO type is reversible and coarsens, and cannot
exhibit the competing-front patterns of
Secs.~\ref{sec:window}--\ref{sec:mechanism}.

The calibrations of Secs.~\ref{sec:flat} and \ref{sec:droplet} are computed under the canonical one-flip dynamics. In both of the geometries that we study, the interface is monotone: the stripe edge only recedes and the droplet only shrinks. So once a site flips, it is absorbed into the growing
majority and each subsequent flip pushes its field further from threshold---it
never revisits the eligibility condition $s_i m_i<0$. The one-flip constraint
therefore never binds, the reversible and irreversible dynamics coincide, and
the calibration is a property of both model variants. Irreversibility becomes active only when
fronts compete, which does not occur for a single eroding stripe or droplet
(Sec.~\ref{sec:mechanism}). 

\section{Flat fronts and stripes: exact lattice pinning}
\label{sec:flat}

The building block is the total kernel weight of a full lattice row at
perpendicular offset $k$ from a focal site. Writing the row sum as
$e^{-k^2/2\sigma^2}\sum_{n\in\mathbb Z}e^{-n^2/2\sigma^2}$ and applying
Poisson summation,
$\sum_ne^{-n^2/2\sigma^2}=\sqrt{2\pi}\,\sigma(1+2e^{-2\pi^2\sigma^2}+\cdots)$,
gives
\begin{equation}
\ell_k=\sqrt{2\pi}\,\sigma\,e^{-k^2/2\sigma^2},
\label{eq:rowsum}
\end{equation}
exact up to a relative correction $2e^{-2\pi^2\sigma^2}<10^{-8}$ for
$\sigma\ge1$. The single-interface results below follow from \eqref{eq:rowsum}
by bookkeeping, and each agrees with direct summation to machine precision.

Consider a stripe of $-1$ spins of width $w$ rows in a $+1$ sea, with a focal
site on the stripe's edge row ($s_i=-1$, so the site flips if $m_i>0$).
Splitting the field by rows, using the evenness of $\ell_k$, and adding the
self-exclusion correction $+1$ ($W_{ii} \equiv 0$) gives
\begin{equation}
m_{\rm edge}(w)=-\big(\sqrt{2\pi}\,\sigma-1\big)
+2\sqrt{2\pi}\,\sigma\sum_{k=w}^{\infty}e^{-k^2/2\sigma^2}.
\label{eq:stripe}
\end{equation}
Two consequences follow. Taking $w\to\infty$ gives the field on a flat
interface,
\begin{equation}
m_{\rm flat}=-\big(\sqrt{2\pi}\,\sigma-1\big)<0\quad(\sigma\ge1),
\label{eq:pinning}
\end{equation}
so a flat interface is strictly stable---pinned---with a pinning constant
available in closed form. The phenomenon is the threshold-dynamics analogue of
propagation failure \cite{keener_propagation_1987}; the content of \eqref{eq:pinning} is the
explicit constant. In the continuum the flat interface is exactly marginal
($m=0$; Sec.~\ref{sec:droplet}), so $-(\sqrt{2\pi}\sigma-1)$ is purely a
discrete lattice effect, growing linearly in $\sigma$ because the missing self-site
is weighed against a row weight $\propto\sigma$.

\begin{figure}[t]
\centering
\includegraphics[width=\columnwidth]{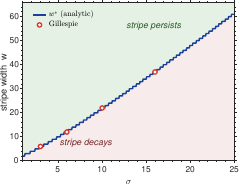}
\caption{Critical stripe width. Solid curve: the exact threshold
$w^{*}(\sigma)$ below which the edge row of a width-$w$ stripe of $-1$ spins in
a $+1$ sea ceases to be eligible, from Eq.~\eqref{eq:stripe}; $w^{*}$ is integer
valued, hence the staircase. Stripes with $w\ge w^{*}$ are pinned and persist
unchanged (shaded region above); narrower stripes erode from both edges to
extinction (below). Circles: the threshold measured from the one-flip Gillespie
dynamics at $\sigma=3,6,10,16$, coinciding with $w^{*}$ in each case. The
transition is sharp because erosion is self-reinforcing (removing a row lowers
$w$ and strengthens the eligibility condition), so a stripe either does not move
at all or vanishes entirely.}
\label{fig:stripe}
\end{figure}

Second, the edge of a finite stripe is eligible if and only if the
monotonically decreasing tail sum in \eqref{eq:stripe} exceeds
$\tfrac12-1/(2\sqrt{2\pi}\sigma)$. This condition defines a critical width $w^{*}(\sigma)$:
stripes with $w\ge w^{*}$ are frozen, and narrower stripes erode from both
edges to extinction via a cascade. The numerically measured survival threshold coincides
with the predicted $w^{*}$ [Fig.~\ref{fig:stripe}]. We note that simulations in this figure and elsewhere in the paper use exact Gillespie dynamics with incremental field updates, a
kernel truncated at radius $\lceil4\sigma\rceil$ (relative truncation error
$<3\times10^{-4}$), and periodic boundary conditions unless stated otherwise.

\section{Droplets: curvature drive, registration statistics, and mobility}
\label{sec:droplet}

\subsection{Ring-mean field and the crossover radius}
\label{sec:ringmean}

Let $\Omega$ be a disk of radius $R$ of $-1$ spins in a $+1$ sea and let
$\bar m(\kappa)$ be the mean of $m$ over the boundary ring, $\kappa=1/R$. To
leading order in curvature\footnote{Up to a $O(\kappa^2)$ correction.},
\begin{equation}
\boxed{
\begin{aligned}
\bar m(\kappa)
={}&-\big(\sqrt{2\pi}\,\sigma-1\big)
+\sqrt{2\pi}\,\sigma^{3}\,\kappa +C(\sigma),\\
&
C(\sigma)\approx0.23\,\sigma .
\end{aligned}
}
\label{eq:ringmean}
\end{equation}
The three terms have different characters, and we state each.

The flat term is exact, Eq.~\eqref{eq:pinning}.

The curvature coefficient is a first-order approximation for small $\kappa$ and
coincides with the continuum threshold-dynamics drive under the same kernel
normalization. Writing $s=1-2\,\mathbf 1_\Omega$ gives
$m=2\pi\sigma^2-2(G*\mathbf 1_\Omega)$ with $G(x)=e^{-|x|^2/2\sigma^2}$ of mass
$2\pi\sigma^2$. For a flat continuum interface $G*\mathbf 1_{\rm half}=\pi\sigma^2$
and $m=0$. Curving the interface to $y=-\tfrac{\kappa}{2}x^2$ and expanding to
first order,
\begin{equation}
\frac{G*\mathbf 1_\Omega}{2\pi\sigma^2}
=\frac12-\frac{\kappa}{2}\int x^2\,\frac{e^{-x^2/2\sigma^2}}{2\pi\sigma^2}\,dx
=\frac12-\frac{\kappa\sigma}{2\sqrt{2\pi}},
\end{equation}
whence $m=\sqrt{2\pi}\,\sigma^3\kappa$. Fitting the measured ring means against
$1/R$ over $R\ge10\sigma$ recovers this coefficient to within $0.6\%$ at
$\sigma=6$ and $10$, anchoring the model to curvature-flow physics.

The offset $C(\sigma)$ is a discrete correction, characterized numerically but
not derived. Its origin is that the boundary ring is a rasterized circle: the
ring \emph{average} of the field over discrete boundary sites differs from the
pointwise flat-interface value by a curvature-independent constant that survives
$R\to\infty$; it is not an artifact of fitting over finite $R$. The residual
$r(R)=\bar m+(\sqrt{2\pi}\sigma-1)-\sqrt{2\pi}\sigma^3/R$ approaches a nonzero
constant rather than decaying---$r\to1.39$ at $\sigma=6$ and $2.29$ at
$\sigma=10$, so $C/\sigma=0.232$ and $0.229$---obtained by subtracting the
exact flat and curvature terms and averaging the residual over
$10\sigma\le R\le500$. Because the exact drive coefficient is used rather than a
fitted one, the estimate is insensitive to slope error, whereas a two-parameter
fit extending to small $R$ is biased by the $O(\kappa^2)$ term. Deriving $C(\sigma)$ is an Euler--Maclaurin problem for the
rasterized-ring average and remains open. Its influence downstream is confined
to the $O(1)$ prefactor of the crossover radius, of which we use only the
scaling. 

The leading-order form describes the ring means for $R\gtrsim10\sigma$; at
smaller radii the $O(\kappa^2)$ term produces a noticeable departure. At the crossover radius itself the expansion
parameter is $\sigma/R_c\sim1/\sigma$, so the leading-order treatment is
self-consistent at large $\sigma$ and marginal for $\sigma\lesssim6$; this is
harmless, since only the $\sigma^2$ scaling of $R_c$ is used. Setting
$\bar m=0$,
\begin{equation}
R_c=\frac{\sqrt{2\pi}\,\sigma^{3}}{(\sqrt{2\pi}\,\sigma-1)-C(\sigma)}
\;\sim\;\sigma^{2},
\label{eq:Rc}
\end{equation}
the scaling following from an $O(\sigma^3)$ numerator over an $O(\sigma)$
denominator, to which $C$ contributes about $10\%$. Droplets with $R\gg R_c$
have $\bar m<0$ and are pinned in the mean; droplets with $R\lesssim R_c$ have
$\bar m>0$. The mean field, however, does not by itself decide spin-flip eligibility, and
we turn next to the fluctuations about it.

\subsection{Two sources of non-uniformity on the boundary ring}
\label{sec:decomp}

Erosion is decided site by site, so what governs it is the \emph{distribution}
of the boundary field, not only its mean. The field on the ring is not uniform,
and it is useful to separate two distinct sources of non-uniformity,
\begin{equation}
m(\theta)=\bar m(\kappa)+a_{\rm aniso}(\theta)+\xi(\theta),
\label{eq:decomp}
\end{equation}
defined operationally by their behavior under sub-lattice translation of the
disk center. Both are deterministic functions of $\theta$ for a given
rasterized disk; they differ in what they depend on.

The term $a_{\rm aniso}(\theta)$ is invariant under sub-lattice translation:
displacing the disk center by a fraction of a lattice spacing leaves it
unchanged. Obtained by averaging $m(\theta)-\bar m$ over sub-lattice center
offsets, it reflects the orientation of the local boundary relative to the
lattice axes. Since $\mathbb Z^2$ is four-fold symmetric, $a_{\rm aniso}$ is
dominated by a $\cos4\theta$ harmonic distinguishing $\langle10\rangle$ from
$\langle11\rangle$ directions. Because the kernel is a smooth Gaussian rather
than a sharp box, this anisotropy is weak, which is why droplets here (under asynchronous dynamics) remain round instead of developing the polygonal Wulff shapes characteristic of
box-neighborhood threshold-growth automata \cite{gravner_random_2006}.

The term $\xi(\theta)$ is the remainder, and it changes completely under
sub-lattice translation. It encodes the registration of the rasterized arc
against the grid, i.e., how each boundary site sits at the sub-lattice level
relative to the ideal circle, and averages to zero over center offsets. For a
given disk, $\xi$ is not random but fully determined. When we refer to its
statistics, we mean the empirical distribution of its values around one ring,
whose spread we write
\begin{equation}
\delta(\sigma)\equiv\mathrm{std}_\theta\,\xi\approx1.26\,\sigma,
\end{equation}
measured to be proportional to $\sigma$ and nearly independent of $R$. That
distribution is platykurtic, and neighboring boundary sites are anticorrelated: a site rasterized inward
tends to be followed by one rasterized outward. Both features matter below---the
first shapes the mobility function, the second delimits what the statistical
description can be used for. None of $\delta$, the distribution shape, or the
correlation is derived here; their derivation from the geometry of rasterized
circles is open.

\subsection{Universal mobility}
\label{sec:mobility}

\begin{figure}[t]
\centering
\includegraphics[width=\columnwidth]{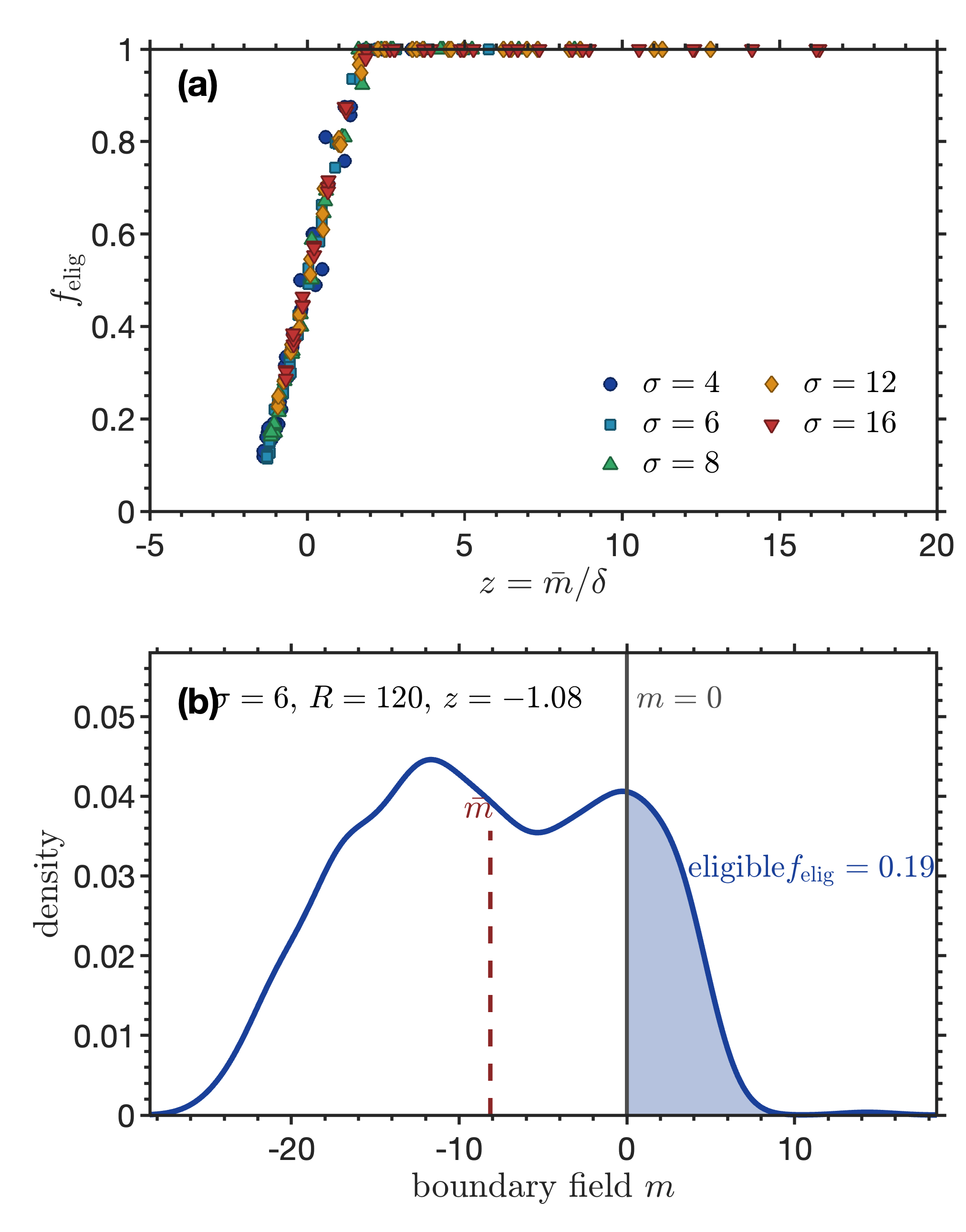}
\caption{Curvature drive and universal mobility.
(a) Fraction of the droplet boundary eligible to flip, versus $z=\bar m/\delta$:
the mean field on the boundary ring, measured in units of the spread of that
field around the ring. Points span $\sigma=4$--$16$ and radii from a few
$\sigma$ to $30\sigma$ and collapse onto a single monotone curve $g(z)$, so the
erosion rate is set by this one ratio rather than by $\sigma$ and $R$
separately.
(b) The origin of the collapse: the measured distribution of the boundary field
$m(\theta)$ for $\sigma=6$, $R=120$ (kernel density estimate over the ring and
over sub-lattice placements of the disk center). A site flips only where $m>0$,
so the eligible fraction is the shaded area, and $z$ measures how far the
distribution's mean (dashed) lies below the threshold in units of its width;
decreasing $R$ shifts the distribution rightward through $m=0$. Here the mean
field is negative, i.e., the ring is pinned on average, yet the positive tail leaves
a fifth of the boundary eligible, the mechanism by which erosion never stalls
and $g(z)>0$ for all $z$ in (a).}
\label{fig:mobility}
\end{figure}

A boundary site is eligible when $m(\theta)>0$, so the eligible fraction of the
ring is a level exceedance of the total deviation $\eta(\theta):=a_{\rm aniso}(\theta)+\xi(\theta)$,
\begin{equation}
f_{\rm elig} = \Pr\nolimits_\theta\!\big[\eta(\theta)>-\bar m\big],
\label{eq:exceedance}
\end{equation}
where $\Pr_\theta$ denotes the fraction of the ring. Two measured properties of
$\eta$ collapse $f_{\rm elig}$ to one variable. First, the spread of $\eta$ is
dominated by $\xi$ and scales as $\delta\propto\sigma$. Second, the distribution
of $\eta$ standardized by $\delta$ is $\sigma$-independent: in lattice units the
rasterized-ring geometry depends only on orientation and on registration phase
within a unit cell, so rescaling by $\sigma$ leaves the standardized statistics
unchanged.\footnote{In lattice units the registration residual $\xi$ has a standardized
distribution that does not depend on $\sigma$: the rasterized-ring geometry
depends only on orientation and registration phase, which are $\sigma$-free, so
rescaling by $\sigma$ collapses all $\sigma$ onto one shape. The weak
orientation dependence ($a_{\rm aniso}$, dominated by $\cos4\theta$) averages
out around the ring and is negligible for the radii used here.} Together these give
\begin{equation}
f_{\rm elig} \approx g(z),\qquad z=\bar m/\delta,
\label{eq:mobility}
\end{equation}
with a single monotone mobility $g$. Here $z$ is formed from the measured ring
mean and spread, not from Eq.~\eqref{eq:ringmean}, so the collapse is unaffected by the
higher-curvature corrections to that expression. The collapse holds to
$\pm0.02$--$0.03$ across $\sigma\in\{4,6,8,12,16\}$ and a decade in $R$
[Fig.~\ref{fig:mobility}], with $g$ rising monotonically from about $0.1$ in the strongly pinned regime
($z\approx-1$), through $0.5$ near $z=0$, reaching $1$ by $z\approx2$
[Fig.~\ref{fig:mobility}(a)]. What organizes the data is the
ratio of the mean drive to the fluctuation scale. We note that $g$ is not the Gaussian $\Phi(z)$: the
platykurtic registration statistics broaden the shoulders, and a Gaussian model
underpredicts eligibility by a factor $1.1$--$1.4$ in the pinned regime.

The statistical description of $\xi$ is used here in a restricted way. By
\eqref{eq:exceedance} the eligible \emph{fraction} depends only on the distribution of $\eta$, so it is insensitive to the anticorrelation noted above. 

Because $g$ is monotone and $\bar m$ increases as $R$ shrinks, the eligible
fraction rises as a droplet erodes: erosion never stalls, every droplet reaches
extinction, and the rate accelerates through $R_c$. Droplets with $R\gg R_c$
($z<0$) still erode, at the slow rate $g(z)>0$ supplied by the registration tail. Throughout, the front remains close to circular and does not develop concave pockets that could arrest it; small
boundary fluctuations stay bounded rather than amplifying (App.~\ref{app:methods}). 

\subsection{Front stability and shape relaxation}
\label{sec:stability}

Under the asynchronous one-flip dynamics, a droplet's boundary is tracked as it erodes through the normalized angular-Fourier amplitudes $a_k/R$. The high modes ($k \geq 4$) stay bounded, so the front does not roughen or finger—it remains smooth. The only mode that grows is $k=2$, a weak large-scale lattice anisotropy that leaves the boundary slightly non-circular wihout introducing fine-scale structure (App.~\ref{app:methods}). Also valuable to note is that, consistent with the model's mechanics, an eroding ellipse retreats fastest at its high-curvature tips and relaxes toward circularity before extinction, which is the
discrete counterpart of curve-shortening rounding
\cite{gage_heat_1986,grayson_heat_1987}. 

\section{Random initial conditions I: the composition window}
\label{sec:window}

\begin{figure}[t]
\centering
\includegraphics[width=\columnwidth]{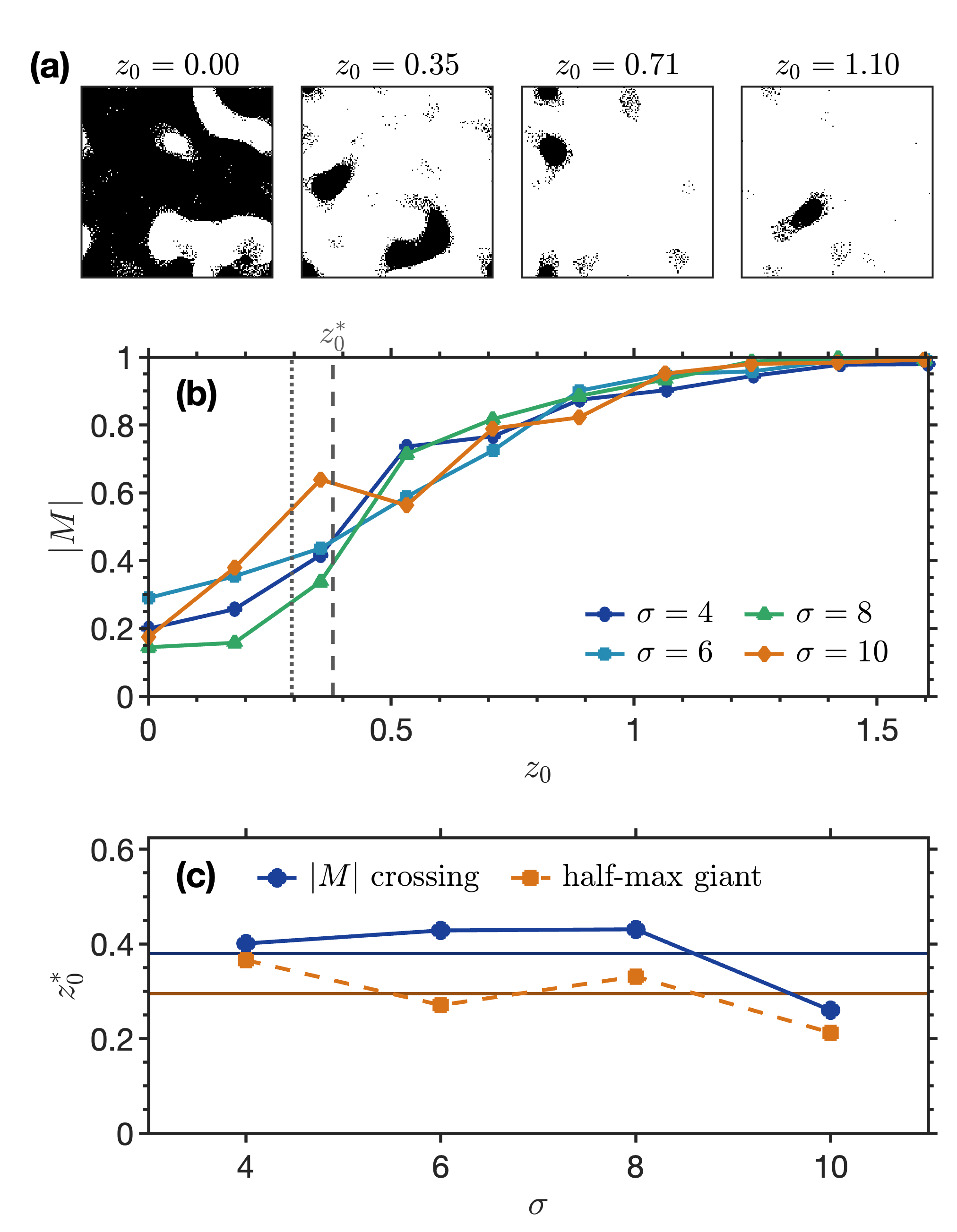}
\caption{The composition window and its control by $z_0$.
(a) Frozen (steady state) configurations at $\sigma=4$ across the window, at increasing
standardized bias $z_0$ (values above each panel). Near $z_0=0$ the state is a
balanced labyrinth. As $z_0$ grows the minority is progressively cascaded away,
and by $z_0\approx1.5$ the state is near-consensus.
(b) The final magnetization $|M|$ versus $z_0$ for $\sigma=4,6,8,10$. Plotted
against $z_0$ the curves for different $\sigma$ collapse, confirming that the
initial condition enters only through the standardized bias; $z_0^{*}$ is marked
under the two operational definitions.
(c) The onset $z_0^{*}$ versus $\sigma$ over $\sigma\in[4,16]$, under both
definitions, with per-$\sigma$ crossing uncertainties from a parametric
bootstrap and horizontal lines at the respective means ($0.36$ and $0.31$). A
weighted regression gives slopes $-0.004\pm0.003$ per unit $\sigma$ for both
definitions, consistent with zero: the cascade boundary sits at a
$\sigma$-independent value of $z_0$ across a fourfold range in interaction
scale.}
\label{fig:window}
\end{figure}

We now take Bernoulli initial conditions, $s_i=+1$ with probability $p_0$
independently. Near balance, the one-flip dynamics freezes labyrinthine patterns
[Fig.~\ref{fig:window}(a)]. Far from balance, it cascades to near-consensus. This
section locates and characterizes the boundary between the two behaviors.

\subsection{An exact control variable}
\label{sec:z0}

At $t=0$ the field \eqref{eq:field} is a weighted sum of i.i.d.\ spins with
\begin{align}
\mathbb E[m] &= (2p_0-1)\big(2\pi\sigma^2-1\big),
\label{eq:mean}\\
\mathrm{Var}[m] &= 4p_0(1-p_0)\big(\pi\sigma^2-1\big),
\label{eq:var}
\end{align}
using $\sum_{j\ne i}\W=2\pi\sigma^2-1$ and $\sum_{j\ne i}\W^2=\pi\sigma^2-1$,
the squared kernel being Gaussian of width $\sigma/\sqrt2$. Both quantities hold up to
Poisson corrections below $10^{-8}$. The standardized initial bias is therefore
\begin{equation}
z_0=\frac{\mathbb E[m]}{\sqrt{\mathrm{Var}[m]}}
\;\xrightarrow[\;\sigma\gg1,\;p_0\to\frac12\;]{}\;
2\sqrt{\pi}\,(2p_0-1)\,\sigma,
\label{eq:z0}
\end{equation}
with $O(1/\sigma^2)$ corrections from the exact forms
\eqref{eq:mean}--\eqref{eq:var}. The initial bias, measured in units of the
initial fluctuations, is thus $(2p_0-1)\sigma$ up to a constant.

The same standardization $z_0$ controls the geometry of the initial field itself. The
covariance of $m$ at separation $r$ is proportional to the kernel
self-convolution, so the correlation function of the standardized field is
$\exp(-r^2/4\sigma^2)$, depending on distance only through $r/\sigma$. In
$\sigma$-rescaled coordinates the standardized initial field is a fixed
unit-variance field with mean $z_0$. Therefore, every sub- or supercritical
excursion-set property of the $t=0$ field is a function of $z_0$ alone. We confirm this directly: at matched $z_0\approx0.35$ the giant-cluster fraction
of $\{m<0\}$ is $\approx0.20$, independent of $\sigma$ within sampling error
($0.19$--$0.20$ for $\sigma=4,6,8$). This is
the static baseline against which the dynamical onset is compared in
Sec.~\ref{sec:dynonset}.

\subsection{Window scaling}
\label{sec:windowlaw}

Let the \emph{composition window} be the interval of $p_0$ over which balanced frozen patterns survive. We find that the window's half-width $\Delta p_0(\sigma)$ obeys a $1/\sigma$ law, whose status
is worth stating precisely, since it combines an exact reduction with one
numerically supported assumption. The exact part is \eqref{eq:z0}: the initial condition enters the
standardized field only through $z_0\propto(2p_0-1)\sigma$. The assumption is
that cascade onset occurs at a fixed value $z_0=z_0^{*}$ independent of
$\sigma$. Granting it,
\begin{equation}
\Delta p_0=\frac{z_0^{*}}{4\sqrt{\pi}\,\sigma}\;\propto\;\frac1\sigma,
\label{eq:window}
\end{equation}
with $O(1/\sigma^2)$ corrections inherited from
\eqref{eq:mean}--\eqref{eq:var}. The assumption is supported numerically in
Sec.~\ref{sec:z0star} but not proved. The support spans $\sigma\in[4,16]$, with
no resolvable drift over a fourfold range in interaction scale, extending well
past the window-scaling fit itself. The natural route to a proof---identifying
onset with a static percolation threshold, which would inherit $z_0$-control
from the static collapse above---is closed by the measurements of
Sec.~\ref{sec:dynonset}.

Since $z_0^{*}$ is $\sigma$-independent (Sec.~\ref{sec:z0star}),
Eq.~\eqref{eq:window} gives $\Delta p_0\propto\sigma^{-1}$ directly. A direct fit
confirms this independently: sweeping raw $p_0$ and locating the onset at each
$\sigma$ gives $\Delta p_0\propto\sigma^{-\alpha}$ with $\alpha=1.05\pm0.09$ over
$\sigma\in[3,8]$, excluding $\sigma^{-2}$—the scaling that raw-mean-field control
(rather than standardized-bias control) would predict. Order parameters vary smoothly through the edge
at all accessible sizes, so the change of behavior is a crossover rather than a
sharp transition. 

\subsection{Location and universality of the boundary}
\label{sec:z0star}

Because the crossover is smooth, $z_0^{*}$ requires an operational definition:
an order parameter and a crossing level. The value obtained depends on that
choice, and we report it rather than average over conventions. Cascade onset defined by the final magnetization crossing $|M|=0.5$ gives
$z_0^{*}=0.36\pm0.03$. Cascade onset defined by the minority giant-cluster fraction
falling to half its balanced value gives $z_0^{*}=0.31\pm0.03$ (means and
scatter across $\sigma\in[4,16]$).
Nearby crossing levels shift these values without affecting the conclusions
below.

The $\sigma$-independence is independent of convention.
Regressing $z_0^{*}$ on $\sigma$ over $\sigma\in[4,16]$ gives slopes $-0.004\pm0.003$ per unit
$\sigma$ for the magnetization definition and $-0.004\pm0.003$ for the
giant-fraction definition (weighted least squares, with per-$\sigma$ crossing
uncertainties from a parametric bootstrap), both consistent with zero. Any
residual drift is bounded by $\lesssim0.09$ in $z_0^{*}$ across the range,
comparable to the run-to-run scatter. Within our resolution the cascade
boundary sits at fixed $z_0$, as assumed in \eqref{eq:window}. The matched-$z_0$ curves collapse to a few percent through the transition
region ($\approx4$--$5\%$ residual spread across $\sigma$, within the bootstrap
sampling-noise floor of $\lesssim5\%$), so the $O(1/\sigma^2)$ narrowing implied
by Eqs.~\eqref{eq:mean}--\eqref{eq:var} is not resolvable in this range. 

\subsection{Irreversibility extends minority survival beyond the static threshold}
\label{sec:dynonset}

The static excursion-set structure of the initial field is controlled by $z_0$
(Sec.~\ref{sec:z0}), so one can ask whether the minority survives to the
absorbing state precisely when some static set of the initial field percolates.
It does not, and the mismatch is the sharpest local expression of what
irreversibility does. 

At balance the field is symmetric, so $\{m<0\}$ covers half the plane, and it sits at the percolation threshold of the initial field. Increasing $z_0$ shrinks this now-minority set below threshold, and its spanning cluster is gone by $z_0 \approx 0.35$. Neither the minority spins themselves, the initially stable minority,
nor the initially eligible minority percolate anywhere in the relevant
range. Yet frozen patterns survive dynamically out to
$z_0^{*}\approx0.3$--$0.4$. There is therefore an interval of composition in
which no static set of the initial field percolates and the minority
nevertheless survives.

The mismatch has a second consequence, for the status of the window law. Because the static geometry is
$z_0$-controlled, a proof that cascade onset occurs at a fixed $z_0^{*}$ would
follow immediately if survival coincided with the percolation of some static set
of the initial field. The mismatch above closes that route for the natural
predictors. We note that we have not excluded the possibility that some finer static
functional coincides with $z_0^{*}$---were one found, the fixed-$z_0^{*}$
boundary would indeed reduce to static percolation---but the ordinary
excursion-set geometry does not supply it. The fixed-$z_0^{*}$ boundary
underlying Eq.~\eqref{eq:window} therefore rests, at present, on the numerical
observation of Sec.~\ref{sec:z0star} rather than on a percolation reduction we
can exhibit.

\section{Random initial conditions II: the frozen morphology}
\label{sec:morph}

\begin{figure}[t]
\centering
\includegraphics[width=\columnwidth]{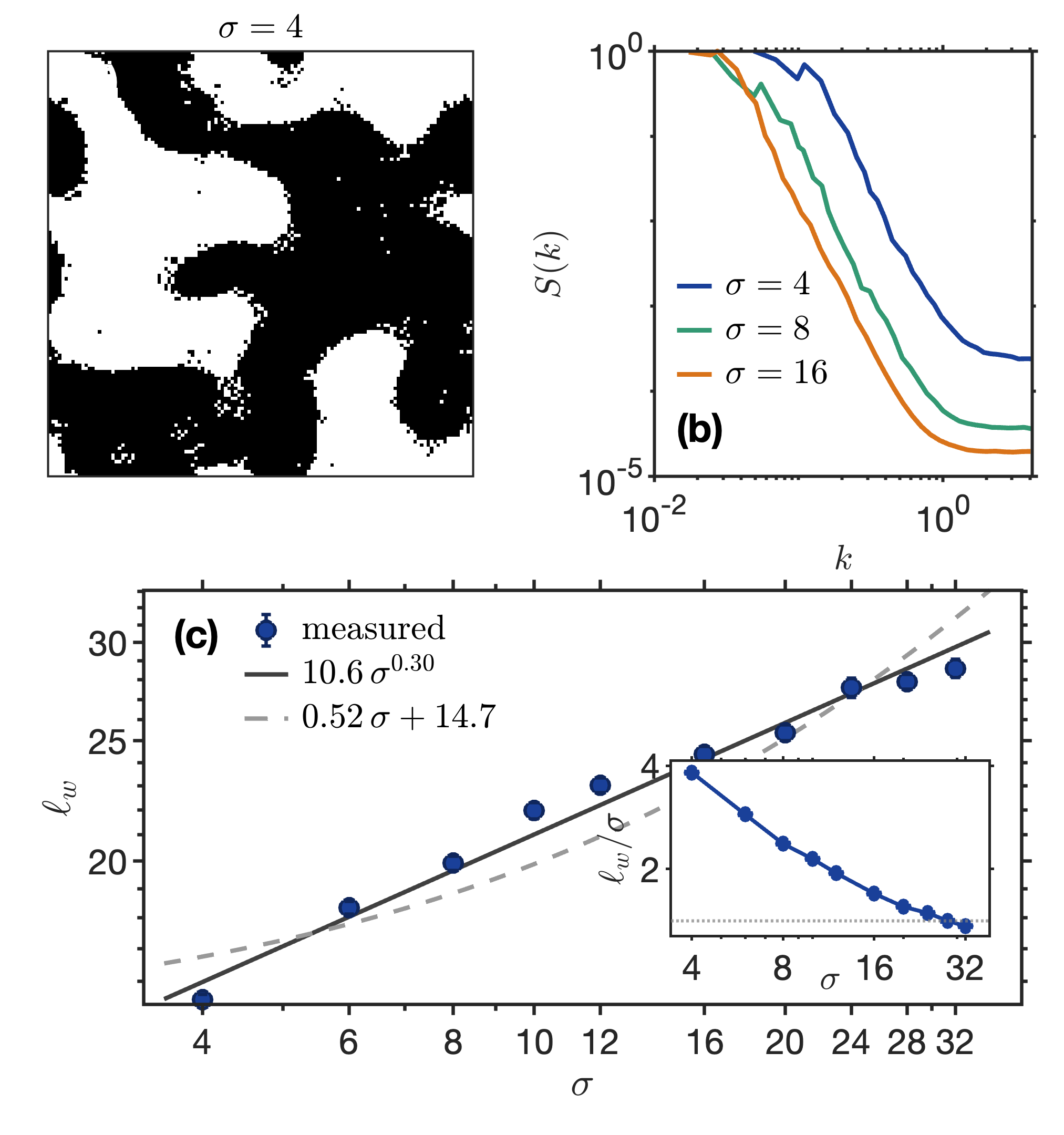}
\caption{Frozen morphology at balance ($p_0=\tfrac12$).
(a) An absorbing-state configuration at $\sigma=4$: a bicontinuous labyrinth.
(b) Radially averaged structure factor at $\sigma=4,8,16$, each normalized to
its maximum. The spectra decay monotonically with no finite-$k$ peak: the purely
positive kernel selects no wavelength, and the pattern is arrested coarsening
rather than a stationary patterned phase.
(c) Feature width $\ell_w=1/w$ versus $\sigma$ (log-log, $48$ realizations per
point at $L=40\sigma$). A power law ($\ell_w\approx10.6\,\sigma^{0.30}$, solid)
describes the data far better than an additive form ($0.52\,\sigma+14.7$,
dashed; rms residuals $0.65$ versus $1.54$), but no single power law fits within
the statistical uncertainties: the effective exponent decreases across the
range, so $\sigma^{\approx0.3}$ is a description over the accessible window
rather than an asymptotic law. Inset: $\ell_w/\sigma$ falls monotonically and
passes below unity near $\sigma=28$, so at the largest scales the frozen feature
is smaller than the interaction range. This excludes $\ell_w\propto\sigma$.}
\label{fig:morph}
\end{figure}

At balance the frozen state is a bicontinuous labyrinth
[Fig.~\ref{fig:morph}(a)], characterized by three measurements.

There is no selected wavelength. The kernel spectrum
$\hat W(k)=2\pi\sigma^2e^{-\sigma^2k^2/2}-1$ is maximal at $k=0$ and decreases
monotonically, crossing zero at a wavelength of order $2\sigma$ into a negative
tail that approaches $-1$ at large $k$. Crucially, it has no positive lobe at
any finite $k$: the real-space kernel is everywhere non-negative, so no
intermediate wavelength is amplified and no Turing-type wavelength selection is
available. The negative short-wavelength tail is stabilizing rather than
destabilizing, consistent with the smoothing at scale $\sigma$ of Sec.~\ref{sec:smooth}. Consistently, the measured
structure factor of the frozen pattern is peakless, rising monotonically as
$k\to0$, the spectral signature of domains at all scales up to $\ell_w$ with no preferred wavelength. 

The frozen state at balance is compact and bicontinuous. Its largest cluster occupies a finite fraction of each phase, not the vanishing fraction a critical percolation state would show. At
the largest accessible size ($L=512$, $\sigma=6$) both phases form spanning
clusters in $94\%$ of realizations, each containing about $82\%$ of its phase's
mass; the fraction of realizations in which both phases span rises with system
size ($56\%$, $69\%$, $94\%$ at $L=128,256,512$), as is consistent with a
bicontinuous morphology whose finite-size cutoff recedes as the box grows. The near-saturation seen here identifies a supercritical, space-filling
structure. It is consistent with the finite characteristic scale $\ell_w$ below,
which a scale-invariant critical state could not possess. The peakless structure
factor separately rules out a wavelength-selected pattern, leaving arrested
coarsening as the description. Our boxes give only
marginal scale separation ($L/\ell_w\sim5$--$30$), so we report the morphology
as compact, bicontinuous, and non-critical rather than attempting a
finite-size-scaling determination of cluster exponents, which we leave to future
work. 

The characteristic feature width grows \emph{sublinearly} with the interaction
range. We measure the wall-density length $\ell_w=1/w$, with $w$ the fraction of
anti-aligned nearest-neighbor bonds; along any line one crosses a wall about
once per feature, so $w\simeq1/\ell$. Over $\sigma\in[4,32]$ and at fixed box ratio $L=40\sigma$ ($48$ realizations per $\sigma$), $\ell_w$ rises from $15.5$ to
$28.6$ while the ratio $\ell_w/\sigma$ falls monotonically from $3.9$ to $0.9$. A smaller $24 \sigma$ control box agrees to within a few percent, with no systematic drift. The
ratio passes below unity near $\sigma=28$: at the largest sizes the frozen
feature is smaller than the interaction range, which excludes any law of the
form $\ell_w\propto\sigma$. A power law describes the data well and much better
than an additive form ($\ell_w\approx10.6\,\sigma^{0.30}$, rms residual $0.64$,
against $0.52\sigma+14.7$, rms $1.54$), with an exponent stable under extension
of the range (it changes from $0.32$ to $0.30$ as the upper limit moves from
$16$ to $32$). We report sublinear growth well described by $\sigma^{\approx0.3}$
over the accessible range; we do not claim this as an asymptotic exponent, but
the monotone decline of $\ell_w/\sigma$ through unity establishes that the
frozen length is not simply proportional to the interaction range. This length
is intrinsic rather than box-set: at fixed $\sigma$, $\ell_w$ converges as
$L/\sigma$ is increased to $96$, so the fixed-ratio protocol above reports the
converged value (App.~\ref{app:methods}).

\section{Irreversibility as the mechanism}
\label{sec:mechanism}

The preceding sections establish what freezes. This section establishes why.
Throughout, \emph{frustrated} denotes a site that has spent its single flip
and nevertheless sits against its field, $s_im_i<0$---a single-flip move that
would lower $H$ but can no longer be taken (Sec.~\ref{sec:model}).

\subsection{Irreversibility, not pinning, freezes the pattern}
\label{subsec:freeze}

\begin{figure}[t]
\centering
\includegraphics[width=\columnwidth]{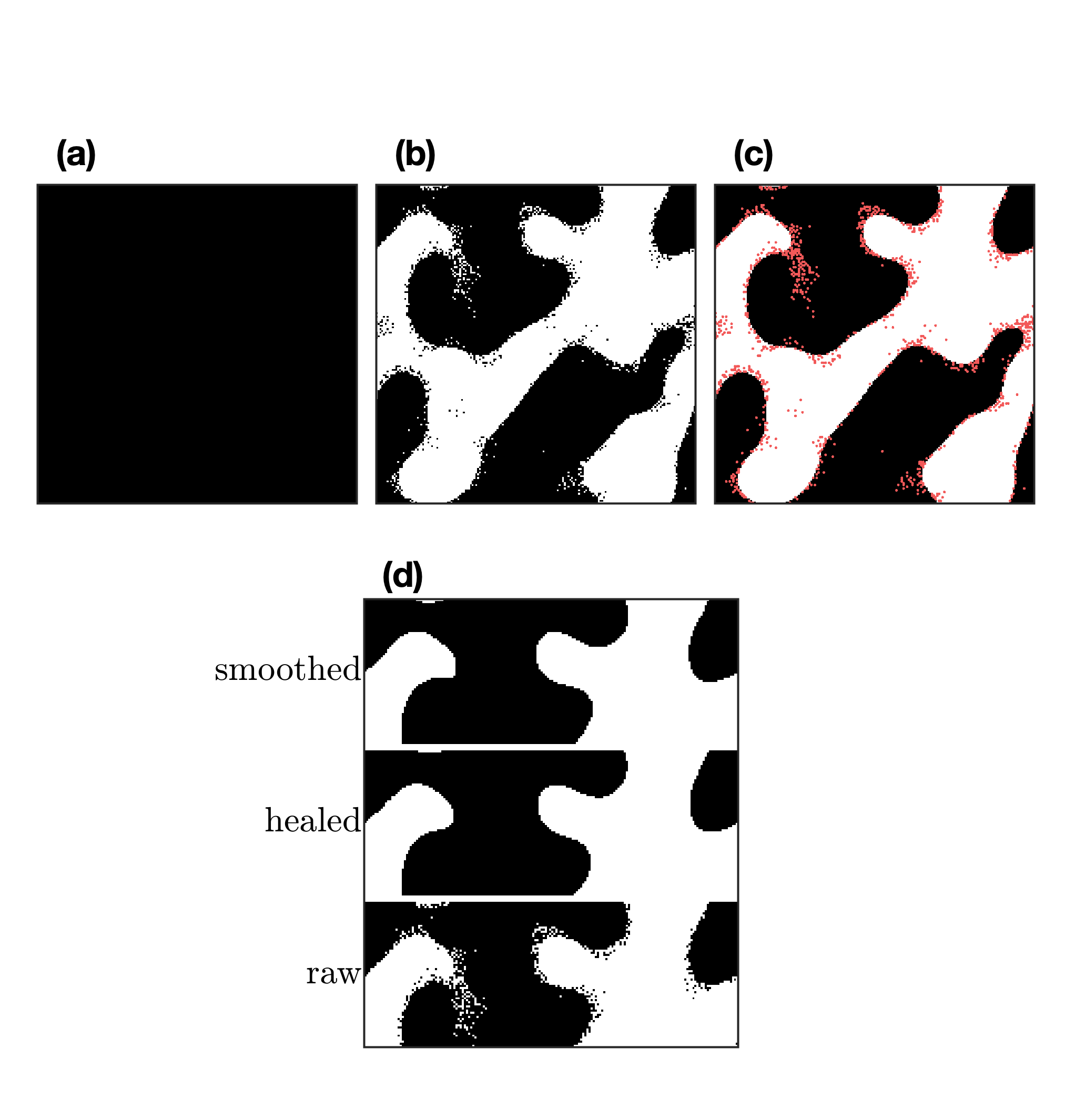}
\caption{Irreversibility, frustration, and the structure of the frozen walls
($\sigma=6$, $L=180$, $p_0=\tfrac12$).
(a) Under reversible dynamics the same initial condition is driven to
near-consensus; (b) under the one-flip rule it freezes into a balanced two-phase
pattern. The only difference between the runs is the irreversibility constraint.
(c) The frozen state with frustrated sites---spins spent against their own local
field---marked in red: they lie almost entirely on the domain walls, tracing the
seams where the competing fronts met. (d) A strip of the frozen state (raw), the
same strip after healing the frustrated sites, and the sign of the
$\sigma$-coarse-grained configuration. Healing removes the sub-$\sigma$ frustrated
roughness while leaving the macroscopic interface unchanged, confirming that
frustration decorates the walls rather than defining them.}
\label{fig:mech}
\end{figure}

From identical balanced initial conditions the two dynamics reach
qualitatively different absorbing states [Fig.~\ref{fig:mech}(a,b)]. The reversible dynamics coarsens away both the fine structure and the balance. Individual runs under reversible dynamics reach either consensus or a few large, straight or gently curved pinned domains—in both cases with high residual magnetization. The one-flip dynamics instead freezes a balanced, bicontinuous labyrinth with low residual magnetization. The difference is not one of degree but of the type of state selected: reversible
descent removes the fine domain structure and rarely arrests, doing so only on nearly
flat interfaces, whereas irreversibility freezes the competing-front pattern
intact. 

Lattice pinning is present in both dynamics and cannot be the origin of the
labyrinth: it acts in the reversible model too, where it produces only the
occasional flat-walled pinned state, never the balanced fine-scale network. That
the reversible model does not always reach full consensus reflects the
single-flip stability of Sec.~\ref{sec:model}---pinning contributes residually
even under reversible descent---but the structure it leaves is coarse and
nearly flat, not labyrinthine. The freezing of a balanced fine-scale pattern is
therefore attributable to the one-flip rule, with pinning a subdominant
correction that acts on both dynamics alike. 

\subsection{Frustration is interfacial}

At the absorbing state frustration is almost entirely a wall phenomenon:
$97$--$98\%$ of frustrated sites lie on domain walls, essentially independent of
$\sigma$. Frustration is thus not a bulk property of the frozen phases but a
decoration of the seams between them---an appreciable fraction of the
interfacial length, but confined to it. The small interior remainder, the
$2$--$3\%$ of frustrated sites off the walls, is a sparse population of point
defects consistent with an origin in sites overrun from several directions at
once.

\subsection{Domain formation, wall placement, and wall locking}
\label{sec:causal}

\subsubsection{Interventions and reference constructions}

We resolve each site by its distance to the nearest coarse domain wall, defined
as the zero level of the frozen configuration coarse-grained at the interaction
scale $\sigma$. Since the frozen state is already smooth at that scale
(Sec.~\ref{sec:smooth}), this coarse-graining alters only the sub-$\sigma$ frustrated skin and
leaves the domain layout intact [Fig.~\ref{fig:dichotomy}]. Two reference
constructions recur below. 

The first reference construction is the \emph{single majority pass}. This operation reassigns every site
once, simultaneously, to the sign of its $\sigma$-Gaussian-weighted field of the initial
condition. It is the pattern a single thresholding of the coarse-grained initial field
would give, with no competition, no advancing fronts, and no spent sites. It
serves as the null against which the dynamics is tested: if the frozen state were
indistinguishable from the smoothed initial condition thresholded once, the irreversible dynamical process would contribute nothing new. For an isolated large feature this null fails trivially---a droplet that the dynamics shrinks to nothing by curvature flow (Sec.~\ref{sec:droplet}) is barely trimmed
by a single pass, and the two could not be confused. The comparison earns its
place here---for the fully developed pattern from random initial conditions, where
it is not evident a priori that iterating the rule adds anything. After all, the smoothed
noise already carries structure at the interaction scale, so one could imagine that a single pass and
our model's frozen state may be similar. Comparing the two thus
separates what the initial field supplies from what our dynamics adds, in the
same spirit as how the comparison with critical percolation in Sec.~\ref{sec:window}
showed that the frozen morphology does not reduce to a percolation. 

The second construction is the \emph{consensus} state: the site-wise majority over many runs from the same initial condition with independent update orders. The purpose of this is to isolate the order-independent skeleton from the order-dependent remainder.

\subsubsection{The three stages}

The frozen pattern is built in three stages: (i) the coarse domains are laid down
by the initial condition and the deterministic dynamics; (ii)
their walls are then placed, predominantly deterministically with a secondary
stochastic contribution; and (iii) frustration locks the placed walls without moving them. 

\begin{figure}[t]
\centering
\includegraphics[width=\columnwidth]{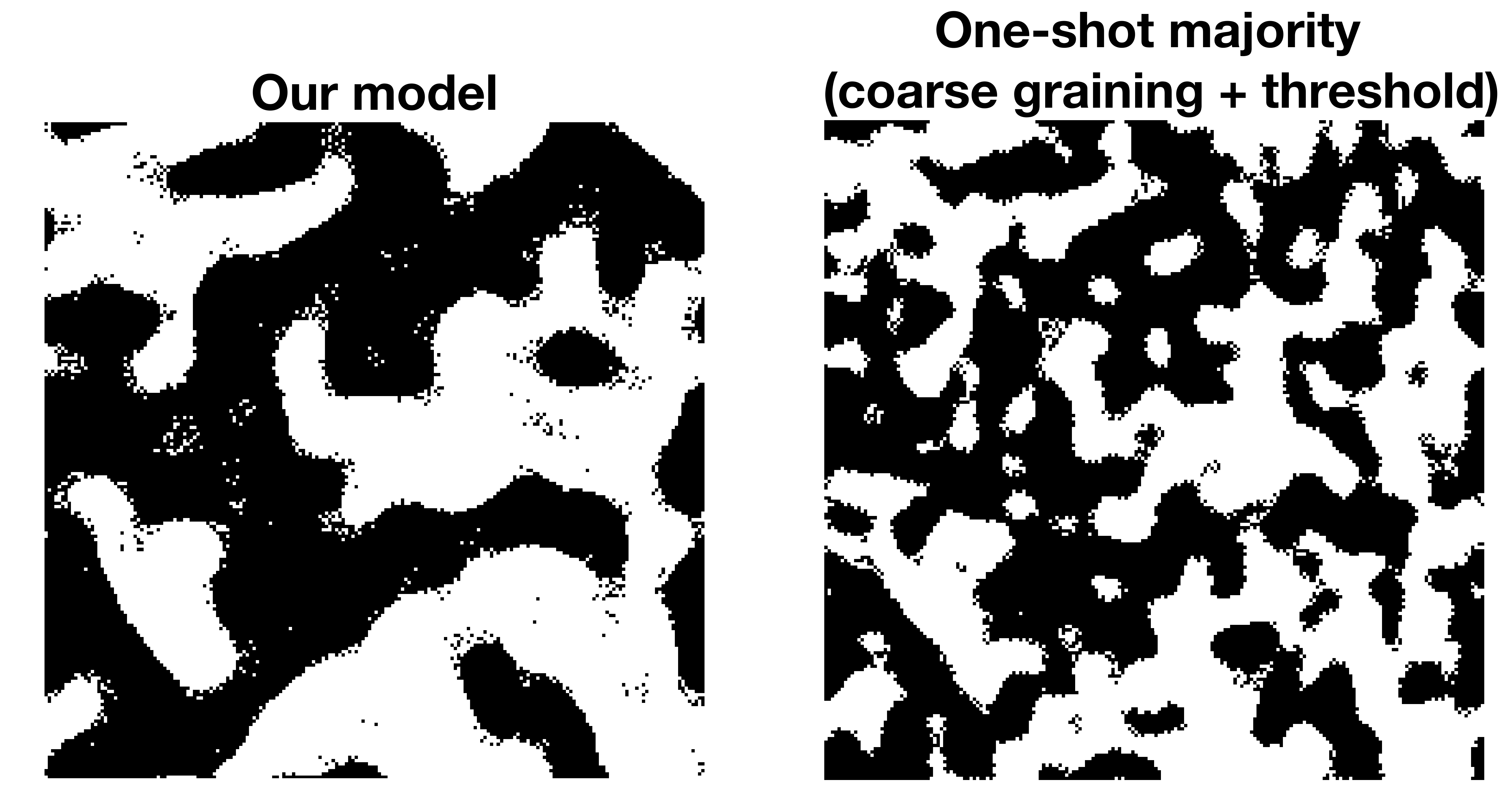}
\caption{The frozen state of our one-flip model (left) and a single majority pass
of the same initial condition (right), at $\sigma=4$, $p_0=\tfrac12$. Both break
the initial noise into two-phase structure at the interaction scale, but the
dynamics does more than threshold the smoothed initial field once: it consolidates
the majority pass's smaller features into markedly fewer and larger domains,
eroding minority pockets and pinching off thin necks. }
\label{fig:maj}
\end{figure}

\begin{figure}[t]
\centering
\includegraphics[width=\columnwidth]{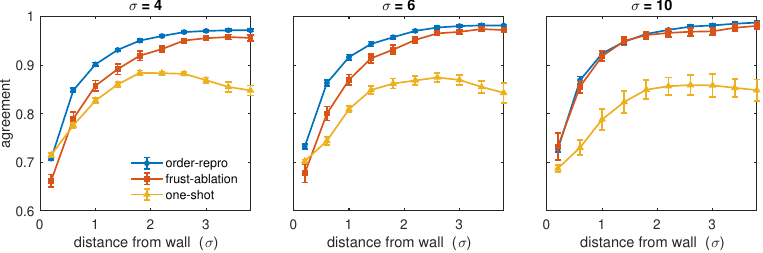}
\caption{Interior/border dichotomy: agreement as a function of distance from the
nearest coarse domain wall (in units of $\sigma$), for $\sigma=4,6,10$ (mean $\pm$
SEM over $8$ initial conditions). \emph{Order-repro:} fraction of update orders
agreeing at a site given the same initial condition. \emph{Frust-ablation:} agreement when the earliest frustrated
sites are exempted from the one-flip rule, against a same-order control.
\emph{One-shot:} agreement with a single majority pass of the initial field. The
order and frustration curves both rise to $\approx0.98$ in the interior and fall
only near the walls: the coarse domains are order-independent and untouched by
frustration ablation, and both effects are confined to near the walls. The one-shot curve saturates near $0.85$, well below the other two.}
\label{fig:dichotomy}
\end{figure}

\emph{Domain formation.} Deep inside a domain, the frozen state is nearly
independent of the update order: the fraction of update orders agreeing at a site
approaches $\approx0.98$ in the interior and falls only near the walls [Fig.~\ref{fig:dichotomy}]. The coarse domains are therefore not
selected by the stochastic dynamics; they are laid down by the initial condition
and the deterministic dynamics acting together. But they are far from what the
the initial field alone implies. Agreement with the single majority pass plateaus near $0.85$ even in the domain interiors, well short of the $\approx0.98$ order-agreement plateau
[Fig.~\ref{fig:dichotomy}], and roughly a quarter of the majority-pass walls have
no counterpart in the frozen state at all. This is the largest way the dynamics
departs from a single pass, and it is visible directly [Fig.~\ref{fig:maj}]: the
interior sites that disagree lie in majority-pass features several times smaller
than the surrounding domains. These are small regions where the smoothed initial field
carries a weak minority sign that a single pass freezes as an island but the
curvature flow of Sec.~\ref{sec:droplet} erodes and absorbs. Although each such
feature is small, removing it reorganizes the layout on large scales: erasing a
minority pocket distorts a domain, and pinching off a thin neck merges the regions it separated. So, the frozen domains are markedly fewer and larger than those of a single pass. The coarse domains are thus the initial condition's majority regions, distorted and consolidated by the dynamics. 

\emph{Wall placement.} Where a wall does survive in both patterns, the dynamics
also moves it---a finer effect than the reorganization above, but a real one.
Taking the single majority pass as reference, whose walls lie at the zero level of
the coarse-grained initial field, the corresponding frozen walls sit about half a
$\sigma$ away, roughly $1.6$ to $2$ times the separation that a change of
coarse-graining width alone produces (App.~\ref{app:methods}). The displacement exceeds this
smoothing null at every scale we examined, though the exact multiple depends on
how the null is defined (this null is a strict one). The walls of domains preserved in both our model and in its null are therefore not the initial contour resmoothed. This displacement is predominantly deterministic: it is present, and of the same size, in the wall of the consensus state construction, so it does not average away over update orders. A smaller, wall-localized stochastic contribution---a first-passage race where opposing fronts meet---rides on top of the deterministic displacement and
grows in relative weight as $\sigma$ decreases, but it is the minority partner. The wall position is set mostly by the deterministic advance of the competing fronts and only fine-tuned by the race. This race is the sole place stochasticity enters the pattern at all---the domains themselves are order-independent [Fig.~\ref{fig:dichotomy}].

\emph{Wall locking.} Frustrated sites lie on the walls, so one might suspect they also play a role in determining the wall positions. They do not. Relaxing the frustrated sites moves only the sub-$\sigma$ wall detail
and leaves the coarse layout intact, and exempting the earliest frustrated sites
from the one-flip rule and rerunning perturbs the coarse layout no more than a
change of update order already does (for any modest fraction of the frustrated
population)---the frustration-ablation and order-agreement curves coincide in the
interior [Fig.~\ref{fig:dichotomy}]. Frustration therefore contributes essentially
nothing to where the walls are placed; its role is to lock them
(Subsec.~\ref{subsec:freeze}). The placement is done while the fronts advance and
spend the contested sites; frustration---the spent sites left against their field
once the fronts have met---is the residue of that spending, and it freezes the
wall where the spending left it. The causal order is placement then locking: the
deterministic dynamics, with a minor stochastic race, sets the domains and their walls; the one-flip
constraint, recorded as frustration, holds them. 

\subsection{The pattern is smooth at scale $\sigma$}
\label{sec:smooth}
The dynamics cannot resolve features finer than the interaction range, since the
field \eqref{eq:field} is a $\sigma$-average and a one-site spike produces the
same field as its $\sigma$-smoothed counterpart. Smoothness of the frozen pattern
is therefore a statement at scale $\sigma$, and at that scale it holds in a
precise operational sense. Healing the frustrated sites---flipping each to agree
with its local field---reduces the interface essentially to the coarse domain
walls of Sec.~\ref{sec:causal}: the healing changes the sign of the
$\sigma$-coarse-grained configuration at no more than $2\%$ of sites, and leaves
the density of coarse domain walls almost unchanged (for example $0.0060$ before
versus $0.0057$ after at $\sigma=16$, at smoothing width $0.7\sigma$). This is
consistent with the wall-localization of Sec.~\ref{sec:causal}. Frustration
therefore carries essentially all of the lattice-scale excess perimeter and almost
none of the macroscopic shape. The frozen labyrinth is a smooth wall network at
scale $\sigma$, placed by the deterministic front dynamics of Sec.~\ref{sec:causal}
and decorated by a sub-resolution frustrated skin. The smoothing is the curvature
flow of Sec.~\ref{sec:droplet}: eligible high-curvature features are eroded during
the invasion, no perturbation mode grows into an instability, and what freezes is
an already-smoothed collision interface.

\section{Discussion}
\label{sec:disc}

The single-interface sector of this model reproduces, with exact lattice
constants, the physics of threshold dynamics on a lattice: curvature-driven
motion with the continuum drive coefficient $\sqrt{2\pi}\sigma^3$, and pinning
with the closed-form constant $-(\sqrt{2\pi}\sigma-1)$, the threshold-dynamics
analogue of propagation failure in discrete reaction--diffusion systems
\cite{keener_propagation_1987}. We regard these results as calibration rather than new
phenomenology.

Irreversibility contributes three things. 1) It supplies a freezing mechanism
distinct from pinning: on initial conditions from which the reversible dynamics
coarsens to consensus, the one-flip rule arrests competing fronts at their
collision seams, freezing the balanced state into a smooth, intricate, and
bicontinuous labyrinth. We witness two interpenetrating domain networks with
structure down to the interaction scale, rather than a global cascade of either
phase or a few large pinned stripes. 2) It admits a quantitative phase structure.
The patterned window is controlled by the exact standardized bias
$z_0\simeq2\sqrt\pi(2p_0-1)\sigma$, with a critical value $z_0^{*}$ independent of
$\sigma$ within resolution, and hence giving a $1/\sigma$ window. This boundary
appears to be genuinely dynamical: it lies beyond every static percolation
threshold of the initial field we examined. And 3) it produces a microscopic
signature with no reversible counterpart: \emph{frustration}. Frustrated sites are
overwhelmingly interfacial, marking the collision seams. They are causally inert
with respect to where the walls sit---they lock the walls rather than place
them---and confined below the interaction scale, so healing them recovers the
$\sigma$-smoothed interface. Together these give a mechanistic rather than merely
descriptive account of irreversible pattern freezing: coarse domains set by the
initial condition and the deterministic dynamics together, which consolidate the
small features a single threshold pass would keep; walls placed off that pass
predominantly by the deterministic advance of the competing fronts, with a
secondary first-passage race fixing only their last fraction of a $\sigma$; and a
sub-resolution frustrated skin that locks the result.

It is worth placing the frozen states precisely against zero-temperature Glauber
dynamics, the reversible relaxation they most resemble. In two dimensions
Glauber reaches consensus except in about a third of runs at balanced initial
conditions, where it freezes into straight, system-spanning stripes held in place
by pinning \cite{spirin_freezing_2001}; the curved bicontinuous labyrinths
found here have no counterpart among its frozen states. In three dimensions
Glauber does produce interpenetrating labyrinthine domains, but those states are
almost never truly static---they contain blinker spins that flip indefinitely at
zero energy cost \cite{spirin_freezing_2001}---whereas the one-flip absorbing states are frozen exactly. The sharpest distinction, however, is not geometric. Because a zero-$T$ Glauber flip
is taken only when it does not raise the energy, no spin ever comes to rest
strictly against its local field; the frozen stripe states are entirely free of
frustration. The one-flip rule removes precisely this guarantee, and the
resulting frustrated density is the observable that most cleanly marks the frozen
states as a product of irreversibility rather than of generic arrested
coarsening. 

Morris \cite{morris_contagion_2000} establishes that coexistent equilibria---configurations with both actions present---can exist near symmetric payoffs for local coordination games, but with no constraint on the relative extent of the two phases and no claim that the dynamics reaches them. Our frozen states are \textit{balanced} (comparable phase fractions across a finite window of initial compositions) and \textit{generic} (the typical absorbing state of the irreversible rule). 

The spatial content of this picture is transferable beyond the present model, and
in two distinct regimes. From disordered initial conditions---the case studied
here---the frozen walls are the seams of a dense domain mosaic: the coarse domains
are laid down by the initial condition and the deterministic dynamics, and their
walls are placed by the curvature-driven advance of the competing fronts, with a
secondary first-passage race fixing only the last fraction of a $\sigma$. This is
the regime of curvature-driven coarsening and grain growth, distinguished here by
the frustration signature that a reversible coarsening would not leave. From
seeded or low-noise initial conditions---a few committed regions invading outward
and meeting---the same walls become isolated competition interfaces between two
irreversibly invading populations, whose location is genuinely a first-passage
outcome, in the spirit of first-passage percolation
\cite{haggstrom_first_1998}. Our results say where such interfaces settle in
both regimes, and that they carry a definite frustration signature, absent
whenever the dynamics is reversible. Wherever two committed invasions compete on a
lattice---rival contagions, competing domains in a solidifying material,
coexisting strategies in a spatial population---the same two observables, the
interface location and its frustrated decoration, are in principle measurable. 

Several questions remain open. Analytically: a proof that the cascade boundary
sits at fixed $z_0$; and first-principles derivations of the registration scale
$\delta\approx1.26\sigma$ and mobility $g$, of the ring offset $C(\sigma)$, and of
why the frozen frustration density is an order-one fraction of the wall density.
Numerically: the feature-width law beyond $\sigma=32$, and the coarse/fine
division of Sec.~\ref{sec:causal} at larger system sizes. The model was motivated
by seeded coordination in crowds. A three-state extension capturing that setting is left to future work. 

\begin{acknowledgments}
This work was conducted using Claude Opus 4.8 (Anthropic, 2026) to help with the literature survey, the implementation of the algorithm, the analysis of the results, and the presentation of the results. The author reviewed and edited the content and takes full responsibility for the content of the published article. 
\end{acknowledgments}

\bibliographystyle{unsrt}
\bibliography{references}
\appendix
\section{Numerical methods}
\label{app:methods}

\emph{Dynamics.} Exact Gillespie simulation on an $L\times L$ torus: the eligible set is maintained
incrementally, an eligible site is chosen uniformly, and time advances by an
exponential variate of rate $n_{\rm elig}$ (we set $\beta=1$). Fields are updated
incrementally over the kernel support, truncated at radius $\lceil4\sigma\rceil$
(truncated mass $<3\times10^{-4}$). Absorbing states are exact (empty eligible
set). All numerics use this one-flip dynamics.

\emph{Ring measurements.} Disks of radius $R$ are rasterized on grids of side
$2\lceil R+6\sigma\rceil+1$, leaving about $6\sigma$ of $+1$ sea on each side. The
field is computed by FFT convolution with the full Gaussian kernel (no spatial
truncation), with the self-exclusion $W_{ii}=0$ handled exactly, so wraparound
contamination is $\sim10^{-8}$. The boundary ring comprises the $-1$ sites
four-adjacent to $+1$; ring means are averaged over the four sub-lattice center
offsets $\{(0,0),(\tfrac12,0),(0,\tfrac12),(\tfrac12,\tfrac12)\}$. The offset
$C(\sigma)$ is obtained by subtracting the exact flat and curvature terms and
averaging the residual over its plateau $10\sigma\le R\le500$, using the exact
slope so the estimate is insensitive to slope error. The registration spread
$\delta$ and the along-ring distribution of $\eta$ are measured on the same rings;
$\delta\approx1.26\,\sigma$ is read from $\mathrm{std}_\theta\,\xi$ after
subtracting the sub-lattice-averaged anisotropy.

\emph{Front stability.} Boundary modes are tracked by rasterizing the boundary
radius $r(\theta)$ on a uniform angular grid and taking its angular Fourier
amplitudes. For the free-erosion test a clean disk is eroded and the normalized
amplitudes $a_k/R$ are recorded at fixed erosion fractions, averaged over
independent runs. The high modes ($k\ge4$) remain small with no unstable band; the
only appreciable low-mode growth is the $k=2$ lattice anisotropy, which does not
roughen the interface. Seeded single-mode boundaries confirm that modes are damped
increasingly fast with $k$ over the resolvable range.

\emph{Universal mobility.} The eligible fraction $f_{\rm elig}$ and the ring mean
and spread $(\bar m,\delta)$ are measured on rasterized disks over
$\sigma\in\{4,6,8,12,16\}$ and radii from $2\sigma$ to $30\sigma$, averaged over
the four center offsets, and each point is placed at $z=\bar m/\delta$ from its
own measured $(\bar m,\delta)$.

\emph{Window measurements.} Sweeps are performed in $x=(2p_0-1)\sigma$ so that
$z_0$ is matched across $\sigma$. Boxes are $L=26\sigma$ for the $z_0^{*}$
determination and $L=32$--$40\sigma$ for the collapse, with $15$ realizations per
point. Crossings are obtained by linear interpolation in $z_0$ of the
run-averaged order parameter, and the drift of $z_0^{*}$ with $\sigma$ is
assessed by ordinary least-squares regression (slope and standard error). The
window exponent is measured independently by sweeping raw $p_0$ (not $x$, which
would impose the $\sigma$-scaling), locating the onset $p_0^{*}(\sigma)$ where
$|M|$ crosses $\tfrac12$, and fitting $\Delta p_0=p_0^{*}-\tfrac12\propto
\sigma^{-\alpha}$ by weighted least squares over $\sigma\in[3,8]$. The
matched-$z_0$ collapse residual is quantified as the median cross-$\sigma$ spread
of the $|M|(z_0)$ curves over the transition region ($0.05<|M|<0.95$), from $48$
realizations per point. Its significance is assessed against a bootstrap noise
floor in which each $\sigma$-curve is recentered to the common $\sigma$-averaged
curve, so the null carries only sampling scatter.

\emph{Morphology.} $\ell_w=1/w$, with $w$ the anti-aligned nearest-neighbor bond
fraction on the torus. Main boxes are $L=40\sigma$ with $16$ realizations, and
$L=24\sigma$ controls. Cluster statistics use four-connectivity with periodic
union--find. Both phases' spanning is tested to assess bicontinuity. Finite-size
extrapolation of $\ell_w$ in $1/L$ was not stable at accessible statistics, so we
report single-large-box values with standard errors and quote the small-box
control as the systematic uncertainty. To confirm that $\ell_w$ measures an
intrinsic scale rather than the box, we fixed $\sigma$ and increased the box
ratio to $L/\sigma=96$. At $\sigma=6$ and $10$, $\ell_w$ shows no systematic
drift beyond $L/\sigma\approx24$, with a residual spread of $1.4\%$ and $3.6\%$
over the largest three ratios---at the level of the sampling error---and the
working $L=40\sigma$ value lies within the converged range.

\emph{Coarse domains and walls.} Throughout the driver analysis, domains are the
sign of the frozen configuration smoothed by a normalized Gaussian of width
$\sigma$ (a coarse-graining, distinct from the self-excluded field kernel of the
dynamics), and the coarse domain wall is its zero level. Because the frozen state
is already smooth at this scale, the coarse-graining changes the domain layout
only within the sub-$\sigma$ frustrated skin. Two reference constructions are
used. The \emph{single majority pass} reassigns every site once to the sign of its
field \eqref{eq:field} of the initial condition---one thresholding of the smoothed
initial field, using the same self-excluded kernel as the dynamics. The
\emph{consensus} state is the site-wise sign of the mean over update orders,
$\mathrm{sgn}\langle S\rangle$; its coarse wall is the fixed coordinate against
which per-site robustness is scored.

\emph{Interior/border dichotomy.} For each of $8$ initial conditions per
$\sigma\in\{4,6,10\}$ ($L=200$, $p_0=\tfrac12$), $M=16$ update orders are run to
the absorbing state and globally sign-aligned to the first. From the per-site
$+1$ fraction $q_i$ we form the order-reproducibility $\max(q_i,1-q_i)$ and the
consensus state $\mathrm{sgn}(2q_i-1)$, whose coarse wall gives each site a
distance $d$ to the nearest wall. Three per-site agreements are binned in $d$ (ten
bins over $0\le d\le4\sigma$, bins with fewer than $20$ sites dropped):
order-reproducibility; frustration-ablation agreement; and agreement with the
single majority pass. The frustration ablation records first-frustration times in
a reference run, exempts the earliest $K=200$ frustrated sites from the one-flip
rule, and reruns \emph{at the same update-order seed as a matched control}, so the
control and ablated runs differ only by the exemption and the comparison isolates
the ablation effect from update-order noise. Curves are aggregated as mean $\pm$
SEM over the $8$ seeds.

\emph{Wall displacement.} The displacement of the frozen walls from the single
majority pass is measured on corresponding walls only: for each majority-pass wall
point we take the distance to the nearest frozen-state wall and retain those
within $2\sigma$, the median of which is the displacement. Points with no frozen
wall within $2\sigma$---about a quarter, lying in regions the dynamics has
reorganized---are not considered wall displacements and are excluded. The null is the same
statistic with the frozen wall replaced by the majority-pass wall resmoothed at a
larger width, capturing the wall motion produced by a change of coarse-graining
alone; the displacement exceeds it by a factor of $1.6$--$2$, and is flat under
variation of the $2\sigma$ correspondence cutoff over $1.5$--$2.5\sigma$. To
separate deterministic placement from the update-order race, the displacement of
the \emph{consensus} wall from the majority-pass wall (deterministic given $S_0$)
is compared with the scatter of individual-order walls about the consensus wall
(the race); the deterministic part dominates at all $\sigma$, the race growing in
relative weight as $\sigma$ decreases.

\emph{Consolidation.} That the interior disagreement with the single majority pass
is small-feature consolidation rather than wall displacement is checked by
this restricting to sites more than $2\sigma$ from the frozen wall and comparing, for
those that disagree with the majority pass versus those that agree, the size of
the majority-pass connected component (four-connectivity, periodic) they occupy:
the disagreeing sites lie in components several times smaller, across $\sigma$.

\emph{Frustration and healing.} A site is frustrated when it is spent and its spin
opposes its field; it is classified as interfacial when four-adjacent to an
opposite spin. Healing flips the frustrated sites to agree with their fields, and
the smoothed interface is $\mathrm{sgn}(G_s*S)$ at smoothing widths $0.7\sigma$
and $1.0\sigma$. That relaxing frustration and exempting a modest fraction of
frustrated sites both leave the coarse domain layout intact---moving only
sub-$\sigma$ detail---is verified against an update-order-only baseline.

\end{document}